\documentclass[letter]{aa}
\usepackage{graphicx}
\usepackage{txfonts}
\newcommand{\psr}{PSR~J1911-1114}

\begin{document}

\title{Discovery of an X-ray nebula in the field of millisecond pulsar PSR~J1911-1114}

\author{Jongsu Lee \inst{1}
	\and
	C. Y. Hui \inst{2}
	\and
	J. Takata \inst{3}
	\and
	Lupin C. C. Lin \inst{4}}

\offprints{J. Lee, \email{skyljs1234@gmail.com}}

\institute{Department of Space Science and Geology, Chungnam National University, Daejeon 34134, Korea \email{skyljs1234@gmail.com}
	\and
	Department of Astronomy and Space Science, Chungnam National University, Daejeon 34134, Korea \email{cyhui@cnu.ac.kr,huichungyue@gmail.com}
	\and
	Institute of Particle physics and Astronomy, Huazhong University of Science and Technology, Chain
	\and
	Department of Physics, UNIST, Ulsan 44919, Korea}

\abstract{We have discovered an extended X-ray feature which is apparently associated with 
millisecond pulsar (MSP) \psr\ from a {\it XMM-Newton} observation, which 
extends for $\sim1^{'}$ and the radio timing position of \psr\ is in the midpoint of the feature.
The orientation of the feature is similar to the proper motion direction of \psr. 
Its X-ray spectrum can be well-modeled by an absorbed power-law with a photon index of
$\Gamma=1.8^{+0.3}_{-0.2}$. 
If this feature is confirmed to be a pulsar wind nebula (PWN), this will be the third case that an X-ray PWN found to be powered by a MSP.  
}
\keywords{stars:pulsars: individual:PSR J1911-1114 --- X-ray:stars --- stars:neutron}

\maketitle

\section{Introduction}

Millisecond pulsars (MSPs) are old neutron stars with a rotation period of $\lesssim20$~ms. 
The population of MSPs has expanded considerably in the last decade (please refer to Hui 2014 and references therein for a review).
In view of the enlarged samples, we have carried out a systematic survey of their X-ray properties with 
all the archival X-ray data accessible through HEASARC Archive\footnote{https://heasarc.gsfc.nasa.gov/docs/archive.html} that can
provide imaging and spectral information (Lee et al. 2018). 47 MSPs are firmly detected in X-ray (Table~1 in Lee et al. 2018), 
X-ray flux upper limits from 36 other MSPs are placed with the existing data (Table~2 in Lee et al. 2018).
In re-examining the X-ray conversion efficiency with these censored data, we have found a relation of
$L_{x}\simeq10^{31.05}\left(\dot{E}/10^{35}\right)^{1.31}$~erg/s (2-10~keV). We have also identified some possible 
different X-ray properties among different black-widow and redback MSPs (cf. Fig.~2 \& 3 in Lee et al. 2018).
 
The archival X-ray imaging data also allow us to search for the existence of diffuse X-ray emission around MSPs. 
We have inspected the archival X-ray images of all 47 X-ray detected MSPs.
In one {\it XMM-Newton} observation, we have identified an interesting extended feature in the field of \psr\ (hereafter J1911).

Currently, there are $\gtrsim70$ PWNe detected in X-ray (cf. Kargaltsev et al. 2013). 
However, only two of them are associated with MSPs, namely the isolated MSP PSR J2124-3358 (Hui \& Becker 2006a; Romani et al. 2017; Hui 2018) 
and the blackwidow PSR B1957+20 (Stappers et al. 2003; Huang et al. 2012). 
This shows that X-ray PWNe associated with MSPs are rare.  

J1911 was discovered in a radio survey with \emph{Parkes} (Lorimer et al. 1996).
It belongs to a binary with an orbital period of $P_{b}\sim2.7$~days with a $\sim0.1M_{\odot}$ helium white dwarf companion.
Its spin period and proper-motion corrected period derivative are 3.6~ms and $1.14\times10^{-20}$~s~s$^{-1}$ respectively,
which imply its spin-down power, surface $B-$field and characteristic age to be
$\dot{E}=9.41\times10^{33}$ erg s$^{-1}$, $B_{\rm surf} = 2.28 \times 10^{8}$ G and $\tau = 4.12$ Gyr respectively.
At a distance inferred from its dispersion measure ($d=1.07$~kpc), its proper motion implies a
traverse velocity of $v_{\rm psr}\sim84$~km/s.
Its position has been found at RA=19$^{\rm h}11^{\rm m}49.283^{\rm s}$, 
Dec=-11$^{\circ}14^{'}22.481^{''}$ (J2000) by radio pulsar timing (Desvignes et al. 2016).

In this Letter, we report the detection of the extended X-ray feature which possibly associated with J1911 and present the results 
from a detailed analysis.

\begin{figure}
\centering
\includegraphics[width=3.0in, angle=0]
{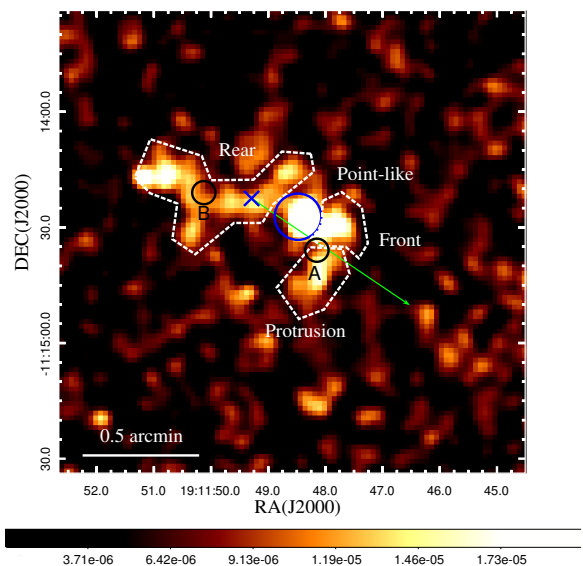}
\caption{Exposure-corrected image of $2'\times2'$ area around PSR J1911-1114 as observed by MOS1/2 in 0.3-10 keV.
Blue cross illustrates the radio timing position of J1911.
Blue circle shows the point-like X-ray source and all the other white dashed regions illustrate
various components of the extended feature.
Green arrow represents J1911's proper motion direction.
The scale bar at the bottom is the pixel values in unit of photon cm$^{-2}$ s$^{-1}$.
Black circles indicate the identified optical sources.
}
\end{figure}

\begin{figure}
\centering
\includegraphics[width=2.8in]{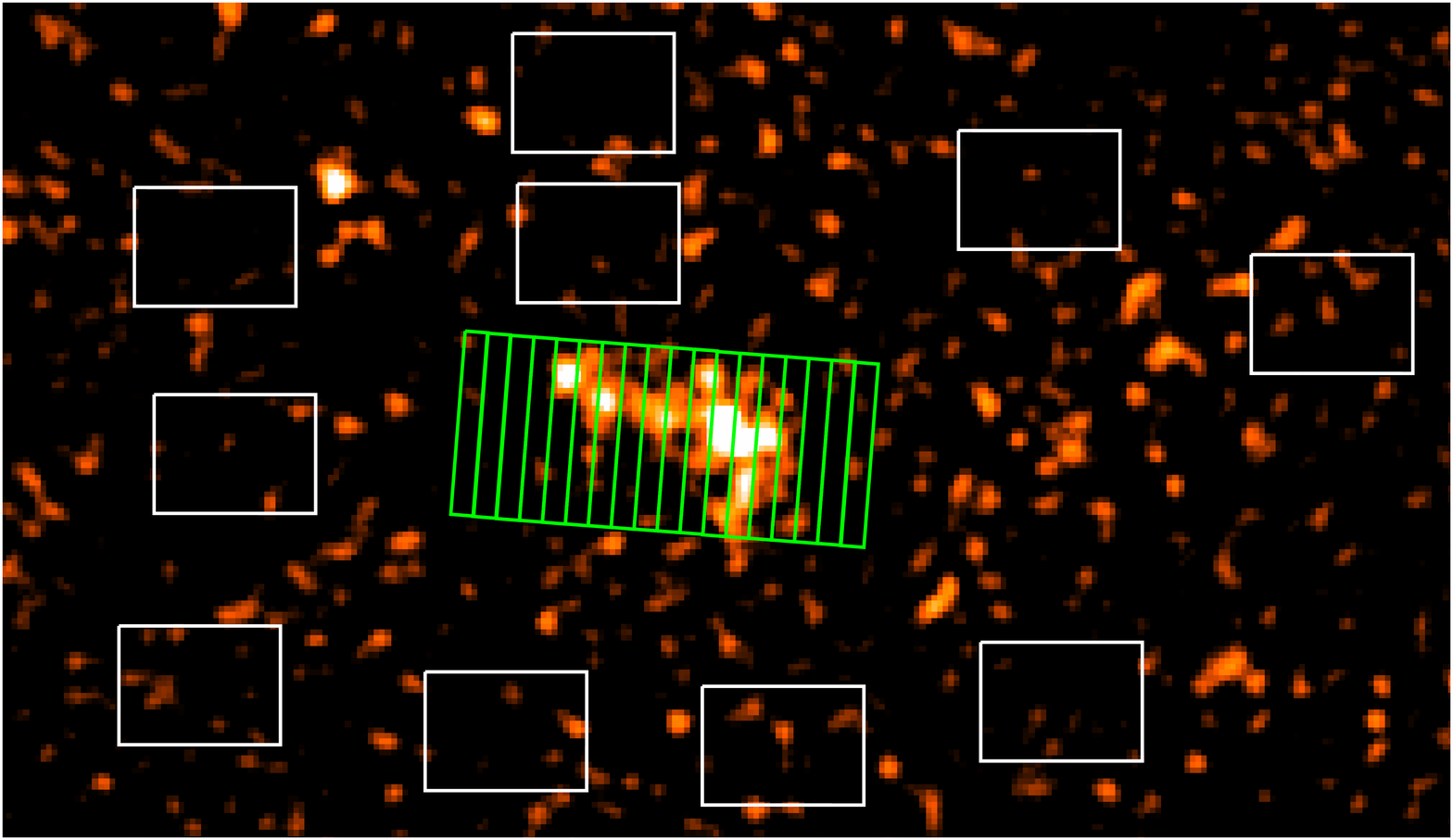}
\includegraphics[width=3.5in]{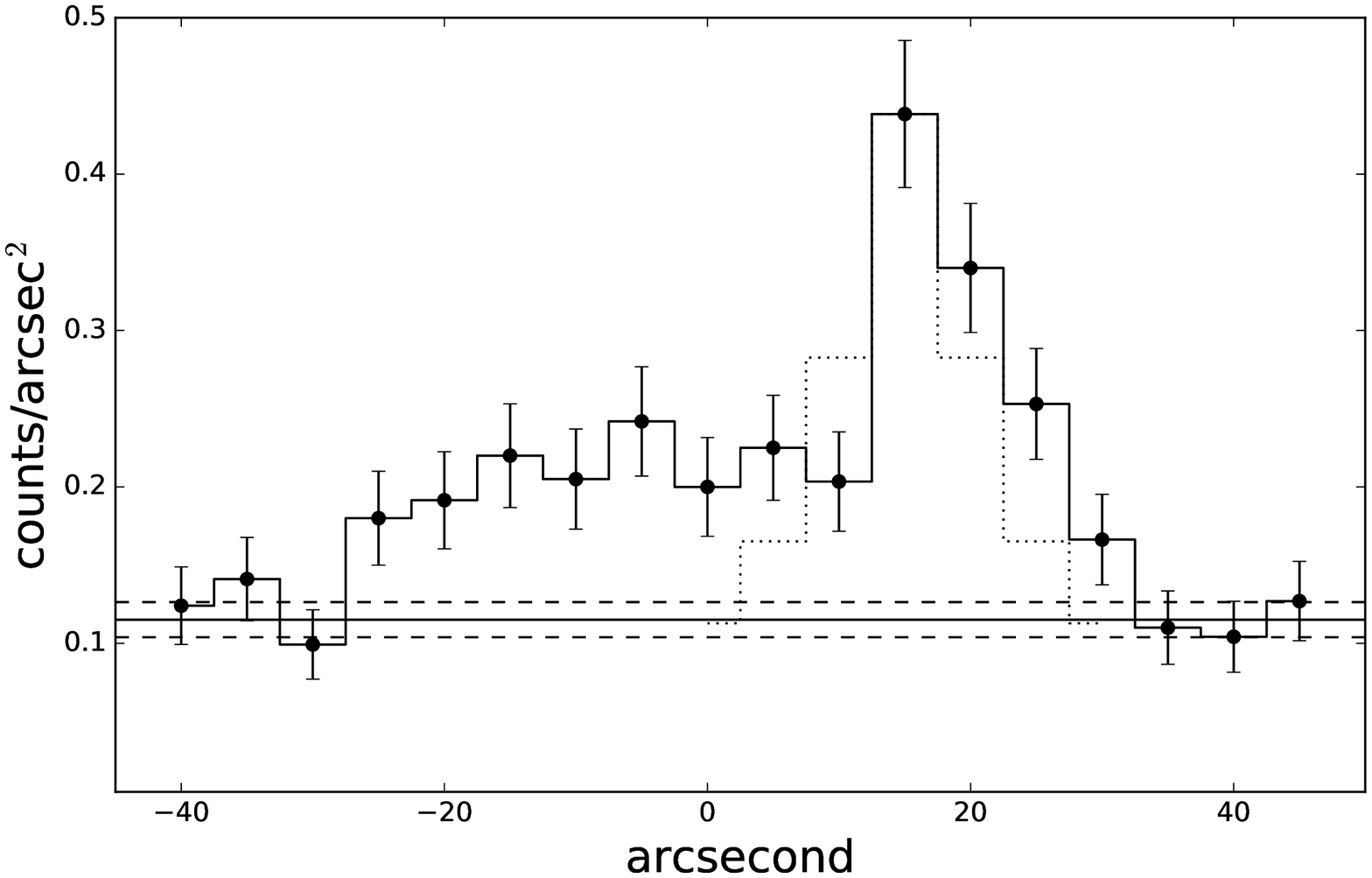}
\caption{
({\it Top panel}) Sampling regions of brightness profile (Green) and background (White). 
({\it Bottom panel}) Brightness profile of the elongated feature.
The solid line and dashed horizontal lines are averaged background level and $1\sigma$ uncertainties.
Zero in the $x-$axis is chosen at the radio timing position.
The profile expected for a point spread function is illustrated by the dotted line.
}
\end{figure}

\section{Observation \& Data Reduction}

J1911 has been observed by {\it XMM-Newton} in 2007-04-27 with a total exposure time of $\sim 68$~ks (ObsID: 0406620201). 
MOS1/2 cameras were operated in full frame mode with a medium filter. On the other hand, PN camera was 
operated in timing mode with thin filter. As the data obtained by PN camera does not provide the 2-D spatial 
information and the 1-D image is too poor to resolve the contribution from J1911, 
we will not consider the PN data in this analysis.

We utilized {\it XMM-Newton Standard Analysis System} (XMMSAS version 17.0.0) software for data reduction and analysis. 
By running the XMMSAS task \texttt{emchain}, we generated the calibrated photon event files for MOS1/2 data with the most updated 
calibration data files. We restricted all the subsequent analyses in an energy range of $0.3-10$~keV.
We further cleaned the data by removing the time segments with the high background. 
In examining the light curves constructed from the full data in each camera, we kept the good time intervals (GTIs) 
with the count rates $<4.5$~cts/s and $<5$~cts/s in MOS1 and MOS2 respectively. After the GTI filtering, effective exposures of $\sim43$~ksec 
are retained in each camera.

\section{Spatial Analysis}

Figure~1 shows a $2^{'}\times2^{'}$  merged MOS1/2 image with a bin-size of $1.25''\times1.25''$ and smoothed with a Gaussian kernel of $3''$. 
A bright extended feature with an orientation similar to the proper motion of J1911 can be clearly seen in this field. 
The whole feature is detected at signal to noise (S/N) ratio of $\sim10\sigma$ in the combined MOS 1/2 data, which was calculated within 40" at the center of radio position (c.f Eq.1 in Hui et al. 2017). 
We have also searched for the objects in the field of the feature with SIMBAD. 
Besides J1911, two optical sources are found in the direction of this feature from 
the United States Naval Observatory (USNO-B1.0) catalog (Monet et al. 2003).
Their locations are illustrated by the black circles A and B in Figure~1. 
These sources are also found in 2MASS catalog (Skrutskie et al. 2006). 
The $B$ band magnitudes of sources A and B as given by the USNO-B1.0 catalog are 15.38 and 16.78 respectively.
To shed light on the nature of these optical sources, we compute their X-ray to optical flux ratio ($F_{X}/F_{\rm opt}$).
We estimated their unabsorbed X-ray fluxes in 0.1-2.4 keV from an aperture with a radius of 4'' centered at their position.
In 0.1-2.4~keV, $\log(F_{X}/F_{\rm opt})$ of A and B are found to be $\sim-3.8$ and $\sim-3.7$ respectively.
These values are consistent with that of F-type star (Krautter et al. 1999).
As there is no known association between a F star and an extended structure in X-ray, 
we conclude that these two objects are unlike to have any association with the X-ray feature. 

\begin{figure}
\centering
\includegraphics[width=3.5in, angle=0]
{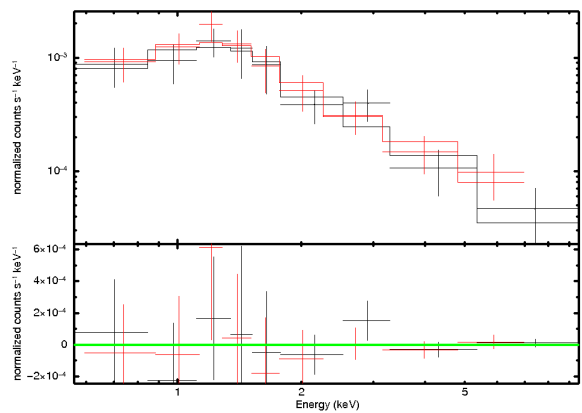}
\caption{Binned spectrum of the whole feature associated with J1911 as observed by MOS1/2 with the best estimated PL model overplotted ({\it Upper panel})
and the fitting residuals ({\it lower panel}).
}
\end{figure}

For investigating the structure of the extended feature, we computed its brightness profile 
by sampling from 18 consecutive bins with a size of $5''\times40''$ that entirely cover the whole feature (see the top panel in Figure ~2). 
For estimating the background level, we randomly sampled ten $35^{''}\times25^{''}$ source-free regions around the feature. 
The background sampling regions are also shown in the top panel in Figure~2.
The brightness profile as obtained from the MOS1/2 data is shown in the lower panel of Figure~2. 
The extent of the feature is estimated by the separation of the points that the profile falls 
into the background in both end. 
Its length is found to be $\sim60^{''}\pm5^{''}$. The uncertainty of the length is estimated by the half width of the adopted bin-size at both ends of 
the feature.

For the peak located at $\sim25^{''}$ from the front end, it has a S/N ratio of $\sim7.0\sigma$. As it resembles a point-like source, 
we have further examine its spatial extent. 
To account for the instrumental effect, we have simulated the point spread function (PSF) at 1.5~keV and the off-axis angle of this point-like 
feature ($\sim16"$) with XMMSAS task {\texttt psfgen}. We then computed the brightness profile from the simulated data in the same way as aforementioned. 
The estimated point source profile is shown in Figure~2 as dotted line. While the profile of the whole feature is significantly different from 
that of a point source, its peak does approximately follow point-like structure locally. 

To further determine the spatial properties of this point-like feature, 
we followed Hui \& Becker (2006b) by using the simulated PSF as a convolution kernal for fitting the unsmoothed image centered 
at this feature with a 2-D Gaussian plus a 2-D constant background model. The best-fit model yields a position of 
RA=19$^{\rm h}$11$^{\rm m}$48.220$^{\rm s}$, Dec=-11$^{\circ}$14$^{'}$27.70$^{''}$ (J2000) with a statistical error of $\sim1.3^{"}$ and 
a full-width half maximum (FWHM) of 5.80$^{+3.16}_{-3.02}$~arcsec. 

Besides the point-like source, we divide the whole feature into several spatial components for the subsequent analysis (see Figure 1). For the long tail-like 
structure trailing J1911's proper motion, we denote it as the `rear' component. It has a S/N ratio of 7.9$\sigma$ in this MOS1/2 data. 
For the emission at the opposition side, we further dissect it into two components, namely the `front' and `protrusion'. Both components are detected S/N$\sim$5$\sigma$. 

\begin{table}
\begin{center}
\scriptsize
\caption{X-ray properties of the whole feature in the field of PSR J1911-1114}
\begin{tabular}{c c c c c}
\hline
\hline
$N_H$ & $\Gamma$/keV& L$_X^{0.3-10}$ & $\chi ^2$ & d.o.f. \\
($10^{21}$ cm$^{-2}$) & & (10$^{30}$ erg s$^{-1}$) \\
\hline
\multicolumn{5}{c}{Power-Law} \\
\hline
\\
$1.70^{+1.31}_{-0.74}$ & $1.79^{+0.32}_{-0.24}$ & $23.48^{+4.13}_{-2.35}$ & 45.15 & 53\\
\\
\hline
\multicolumn{5}{c}{Thermal Bremsstrahlung} \\
\hline
\\
$0.605^{+0.0.820}_{-0.380}$ & $6.75^{+27.2}_{-2.20}$ & $25.07^{+2.80}_{-3.29}$  & 42.64 & 53 \\
\\
\hline
\hline
\end{tabular}
\end{center}
\footnotesize
Note. Uncertainties are computed from the 68.2\% credible intervals of the posterior distributions estimated through MCMC.
$L_x$ based on absorption corrected flux and are adopted distance of 1.07 kpc.
\end{table}

\section{Spectral analysis}

\begin{figure}
\centering
\includegraphics[width=3.5in]
{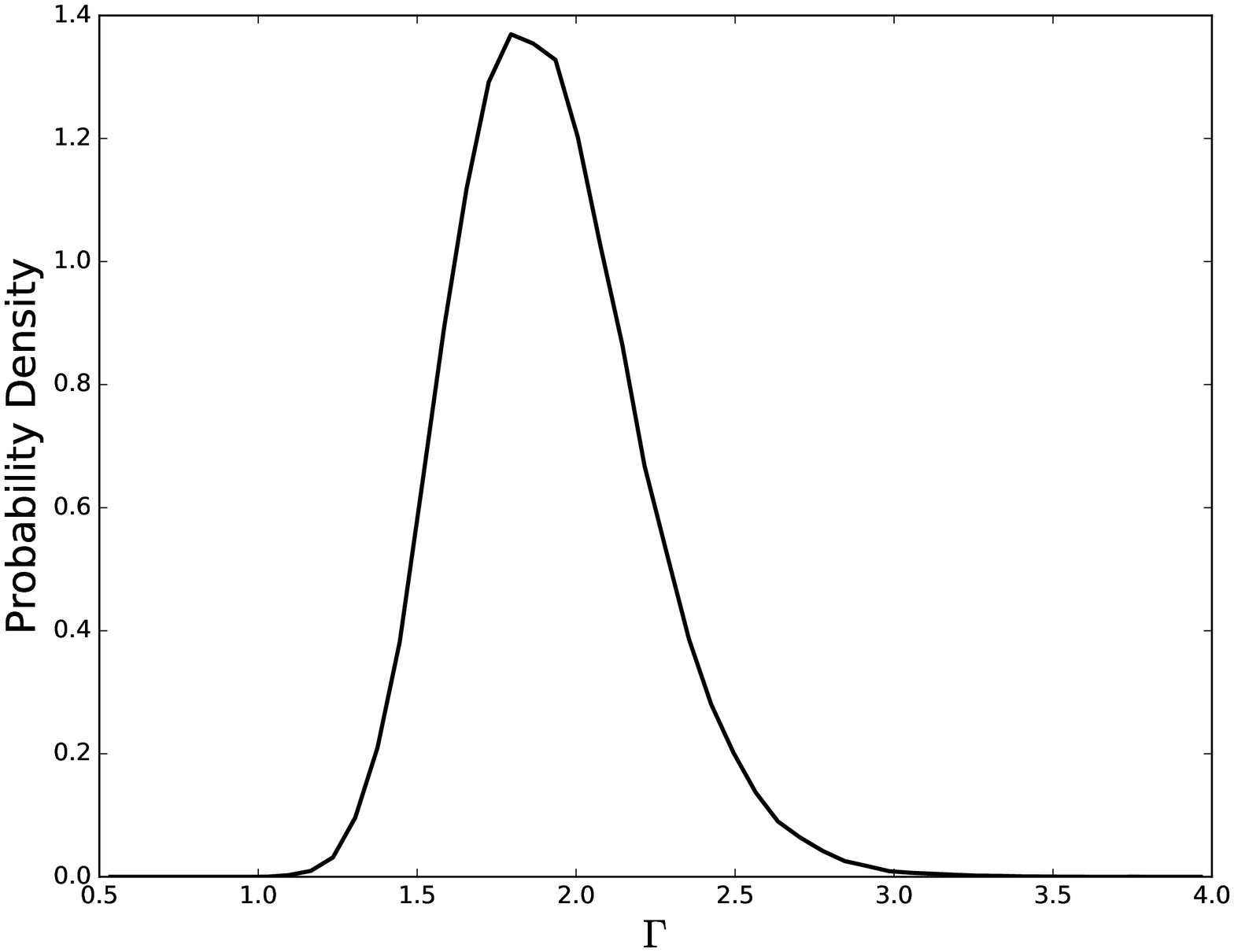}
\includegraphics[width=3.5in]
{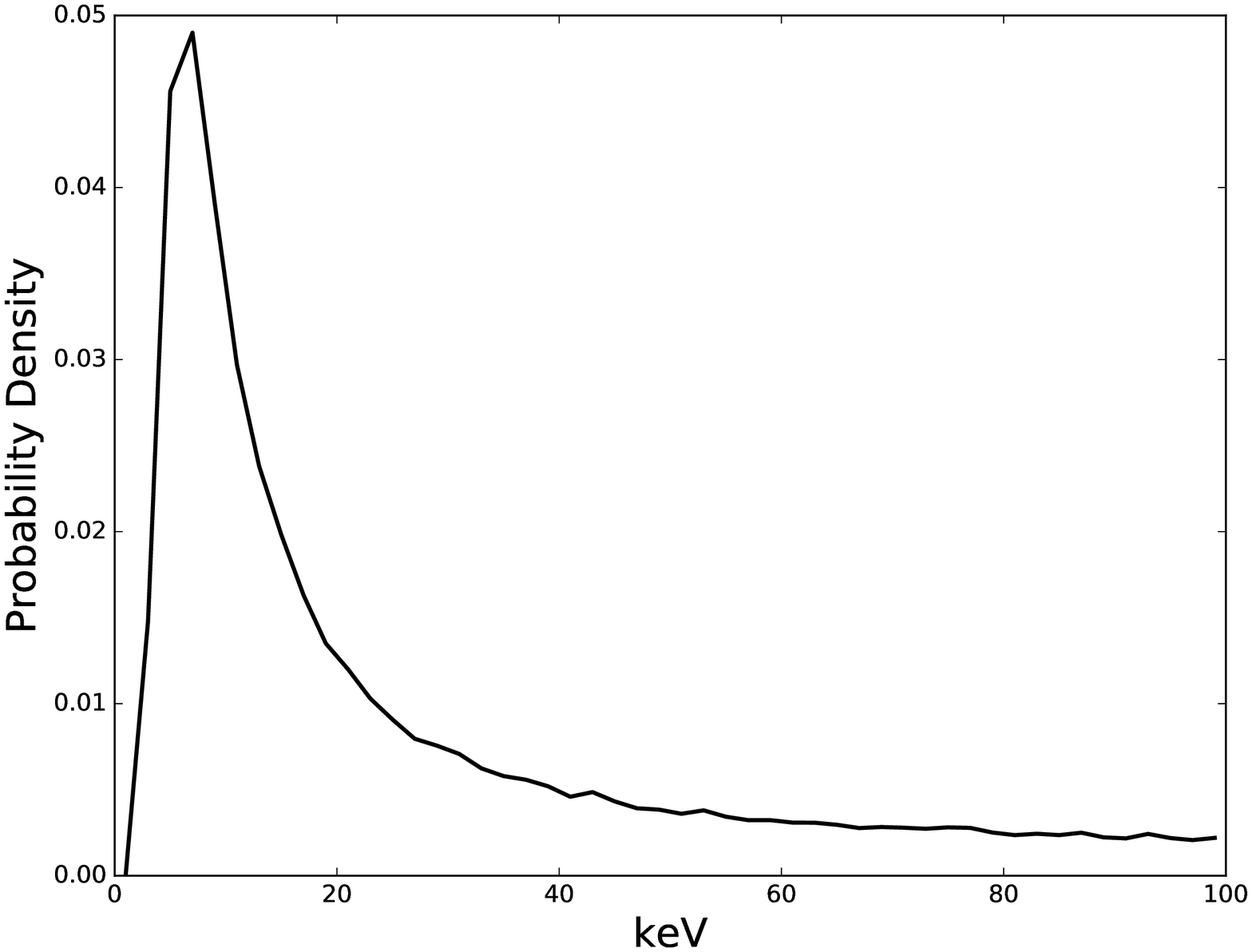}
\caption{1D marginalized posterior probability distributions of photon index ({\it Upper panel}) and the plasma temperature 
({\it lower panel}) as inferred from the non-thermal and thermal scenarios respectively.}
\end{figure}

We extracted the MOS1 and MOS2 spectra of the whole feature from all the regions shown in Figure~1. 
Background spectra are sampled from the nearby source-free regions in the corresponding cameras. 
The background-subtracted counts of the feature collected by both MOS1 and MOS2 cameras is $\sim$270~cts. 

With an aim to tightly constrain the spectral parameters, we fit the spectra obtained by MOS1 and MOS2 simultaneously with 
an absorbed power-law (PL) by using the {\it XSPEC} package (ver.12.9).
For the parameter estimation, assuming uniform priors for all parameters, 
we obtained a multi-dimensional posterior probability distribution for all parameters by Markov Chain Monte Carlo (MCMC). 
We adopted the sampling algorithm devised by Goodman \& Weare (2010).
After initial burn-in, we used a set of 8 walkers and each goes for 50000 steps (cf. Danilenko et al. 2015). 
Using these 400000 samples to approximate the posterior probability distribution (cf. Figure~4), 
we estimated the best-fit and the 68.2\% credible intervals of the model parameters through the bayesian approach. 

The whole feature can be well-described by an absorbed PL model with a column density 
of $N_{H}=1.70^{+1.31}_{-0.74}\times10^{21}$~cm$^{-2}$ and a photon index of $\Gamma=1.79^{+0.32}_{-0.24}$.  
The inferred $N_{H}$ is comparable to the total Galactic HI column density in the direction of J1911 (Kalberla et al. 2005; Dickey \& Lockman 1990).
The unabsorbed flux in 0.3-10~keV is found to be $f_{x}=1.71^{+0.30}_{-0.17}\times10^{-13}$~erg~cm$^{-2}$~s$^{-1}$. 
The best-fit PL and the residuals for the observed spectra of the whole feature are shown in Figure~3. 

We have also considered a purely thermal scenario by fitting the X-ray spectrum of J1911 with an absorbed thermal bremsstrahlung 
model (cf. Marelli et al. 2013). The best-fit model yields an $N_{H}=6.05^{+8.20}_{-0.38}\times10^{20}$~cm$^{-2}$ and 
a plasma temperature of $kT=6.75^{+27.2}_{-2.20}$~keV. The unabsorbed flux is found to be $f_{x}=1.83^{+0.20}_{-0.24}\times10^{-13}$~erg~cm$^{-2}$~s$^{-1}$ 
(0.3-10~keV). In examining the fitting residuals, we do not identify any systematic 
deviation. However, in comparing the posterior probability distribution of $kT$ and that of $\Gamma$,
we found that $kT$ is poorly constrained. 

We have also examined if there is any spectral variation among different spatial components as identified in Figure~1. 
However, within the tolerance of the statistical errors, we do not find any significant variation across the whole feature.

\section{Summary \& Discussion}

We have discovered an $\sim60^{"}\pm5^{"}$ extended X-ray feature in the field of the MSP J1911 at a significance of $\sim10\sigma$. 
At a distance of $d\sim1$~kpc, this corresponds to a physical size of $\sim10^{18}$~cm.
Its X-ray spectrum can be well-described by a PL with $\Gamma\sim 1.8$. 
On the other hand, a thermal bremsstrahlung with $kT\sim6.8$~keV can result in a comparable goodness-of-fit though the parameter 
cannot be properly constrained. In the following, 
we discuss the implications of both scenario.

Comparing to the other two MSPs (i.e. PSRs B1957+20 and J2124-3358) have X-ray nebulae associated, the X-ray morphology of J1911's nebula 
has two major differences. First, while the X-ray tails powered by PSRs B1957+20 and J2124-3358 have a physical 
length at the order of $\sim3\times10^{17}$~cm 
(cf. Stapper et al. 2003; Hui 2008), the feature associated with J1911 is $\sim10^{18}$~cm which is about three times longer. 
Second, the pulsars are located at the end of the 
tails in the cases of PSRs B1957+20 and J2124-3358. But J1911 is located around the middle of the feature. 
For PSRs B1957+20 and J2124-3358, the morphologies of their X-ray nebulae suggest an origin of a synchrotron nebula 
resulting from the postshock flow. On the other hand, the aforementioned differences between the case of J1911 and these two MSPs 
suggest a different nature of its nebula. 

Since J1911 is at midpoint of the feature (cf. Figure~2), this prompts us to speculate whether the feature is a result of 
bipolar outflow along the spin axis of the pulsar. 3D relativistic magnetohydrodynamic simulations of PWN 
have demonstrated the formation of the bipolar outflow (cf. Figure~14 in Porth et al. 2013). The size of the simulated outflow is found to be 
comparable with that of J1911 nebula. Also, assuming it is indeed a bipolar outflow, it suggests the spin axis of J1911 can be in parallel with 
its proper motion (cf. Figure~1). This is consistent with the spin-velocity alignment that has been found in many other pulsars. (Noutsos et al. 2012). 
The brighter part of the feature in the forward direction can be a result of the ram pressure exerted by the motion of J1911.  

Assuming a non-thermal scenario, we further model the feature with the one-zone PWN model (Cheng et al. 2006; Chevalier 2000). The model luminosity 
is given by: 
$$L(0.3-10keV)={1\over 2}({p-2\over p-1})^{p-1}({2\over 2-p})
({6e^2\over 4\pi^2 m_e c^3})^{(p-2)/4}$$
$$\times\epsilon_e^{p-1}\epsilon_B^{(p-2)/4}\gamma_w^{p-2}R_s^{-(p-2)/2}{\dot E}^{(p+2)/4}
[{\nu^{1-p/2}_{10keV} - \nu^{1-p/2}_{0.3keV}}].$$

In fast-cooling regime, the observed photon index of the extended feature implies a spectral index of 
$p\sim2.2$ for the distribution of the synchrotron emitting electrons. 
The shock radius is estimated by $R_s\sim1.8\times 10^{16}{\dot E}_{34}^{1/2}n^{-1/2}v_{p,100}^{-1} {\rm cm}$ ($\sim$0.02" at 1~kpc),
where the spin down power, $\dot E_{34}$ is in units of $10^{34}$ ergs s$^{-1}$,
the number density, $n$ of the interstellar medium is in units of 1 cm$^{-3}$,
and the pulsar velocity, $v_{p,100}$ is in units of 100 km s$^{-1}$.

Following Cheng et al. (2006), we assume a fractional energy density of electrons $\epsilon_e=0.5$,  
a fractional energy density of the magnetic field $\epsilon_B$=0.003 and a Lorentz factor of the wind particles $\gamma_w$=10$^{6}$ to be the prior inputs.
With these physics inputs for a typical PWN, we attempted to compute the model X-ray luminosity by MCMC. 
For $p$ and $R_{s}$, we fixed their values inferred from the observed properties. 
$\epsilon_e$, $\epsilon_B$ and $\gamma_w$ were taken as the free parameters and assume their former to be Gaussians with the mean and standard deviation
of the aforementioned values. 
Using 40000 samples, the credible interval of the model luminosity in 0.3-10~keV was calculated within 68.2\% with $\texttt{pymc3}$.\footnote{https://docs.pymc.io/}
It is found to be $(1.3\pm0.1)\times10^{31}$ erg/s. The posterior distributions of $\epsilon_e$, $\epsilon_B$ and $\gamma_w$ are shown in Figure~5. 

Comparing the model PWN luminosity with the observed $L_{x}$ (i.e. $\sim2\times10^{31}$~erg/s), they differ by a factor of $\sim$2. 
We would like to point out that such discrepancy can be reconciled by the uncertainty of distance. 
The observed $L_{x}$ is based on the distance inferred from the radio dispersion measure of J1911 $d_{\rm DM}=1.07$~kpc. 
Lee et al. (2018) have shown that the uncertainty of dispersion measure inferred distance is typically $\sim40\%$. 
As $L_{x}\propto d^{2}$, this adds a factor of $\sim2$ to the error budget of $L_{x}$. 
Therefore, we cannot exclude the possibility of a PWN origin for the extended X-ray feature of J1911. 
For better constraining its distance and hence its energy flux, parallax observations are encouraged. 

\begin{figure}
\centering
\includegraphics[width=3.0in, height=3.0in]
{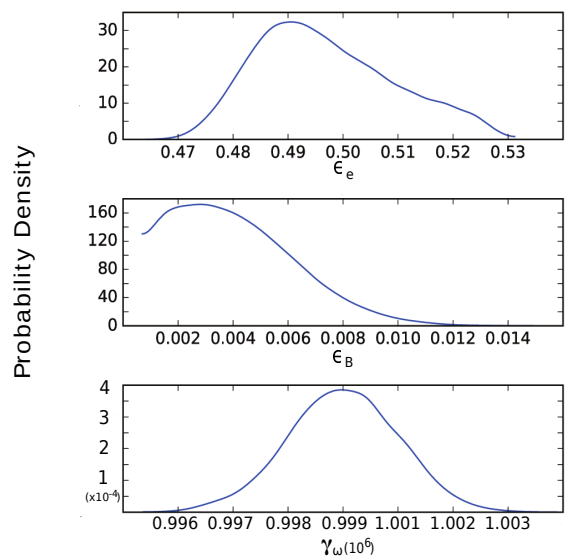}
\caption{The posterior probability distributions for the free parameters $\epsilon_e$, $\epsilon_B$ and $\gamma_w$. 
}
\end{figure}

Besides the PL model, we found that the X-ray spectrum of the J1911 feature can also be described by a thermal bremsstrahlung model. This might suggest 
a mechanism similar to that proposed for explaining the properties of the tail of Morla (i.e. PSR~J0357+3205; Marelli et al. 2013). In such scenario, the 
X-ray emission of the feature is originated from the shocked ISM (Marelli et al. 2013). Assuming a strong shock and an adiabatic index of 5/3, 
the plasma temperature can be estimated as $kT\simeq 3\mu m_{p} v_{\rm p}^{2}/16$, where $m_{p}$ and $\mu$ are the proton mass and the molecular weight 
of the ISM. According to this, the pulsar velocity of $v_{p}\sim84$~km/s implies the temperature of the shock-heated plasma at the order 
of $kT\sim0.01\mu$~keV. It appears that the difference between this estimate and that inferred from the thermal bremsstrahlung fit (i.e. $kT\sim6.8$~keV) 
cannot be reconciled by any reasonable choice of $\mu$. In view of this, we do not favor the thermal bremsstrahlung model to describe the X-ray emission
of the J1911 feature. 

\begin{acknowledgements}
	JL is supported by BK21 plus Chungnam National University and the National Research Foundation of Korea grant 2016R1A5A1013277;
	CYH and LCCL are supported by the National Research Foundation of Korea grant 2016R1A5A1013277;
        JT is supported by the NSFC grants of China under 11573010, U1631103 and 11661161010.
\end{acknowledgements}

\end{document}